\documentclass[12pt]{amsart}
\usepackage{amscd}
\newtheorem{theorem}{Theorem}[section]
\newtheorem{lemma}[theorem]{Lemma}

\theoremstyle{definition}

\newtheorem{example}[theorem]{Example}
\newtheorem{remark}[theorem]{Remark}
\numberwithin{equation}{section}
\begin{document}

\def\CO{\mathbb C}
\def\RE{\mathbb R}
\def\C{\mathbb C}
\def\E{{E}}
\def\Q{{Q}}
\def\fh{{\mathfrak h}}
\def\p{\par\noindent}
\def\S{\mathcal S}
\def\K{\mathcal K}
\def\H{\mathcal H}
\def\A{\mathcal D}
\def\F{\mathcal F}
\def\I{I}


\title[Lax-Phillips scattering for generalized Lamb
models]{Dynamics and Lax-Phillips scattering for  generalized Lamb models}

\author{Massimo Bertini}

\address{Dipartimento di Matematica, Universit\`a di Milano, I-20133
Milano, Italy}

\email{bertini@mat.unimi.it}

\author{Diego Noja}

\address{Dipartimento di Matematica e Applicazioni, Universit\`a di
Milano-Bicocca, I-20126, Milano, Italy}

\email{diego.noja@unimib.it}

\author{Andrea Posilicano}

\address{Dipartimento di Fisica e Matematica, Universit\`a dell'Insubria, I-22100
Como, Italy}

\email{posilicano@uninsubria.it}

\begin{abstract} 
This paper treats the dynamics and scattering of a model of coupled 
oscillating systems, a finite dimensional one and a wave field on the 
half line. The coupling is realized producing the family of
selfadjoint extensions of the suitably restricted self-adjoint
operator describing the uncoupled dynamics. The spectral theory of the 
family is studied and the associated quadratic forms constructed. 
The dynamics turns out to be
Hamiltonian and the Hamiltonian is described, including the case in
which the finite dimensional systems comprises nonlinear oscillators;
in this case the dynamics is shown to exist as well. In the linear
case the system is equivalent, on a dense subspace, 
to a wave equation on the half line with 
higher order boundary conditions, described by a differential
polynomial $p(\partial_x)$ 
explicitely related to the model parameters. In terms of such
structure the Lax-Phillips scattering of the system is studied. In
particular we determine the scattering operator, which turns out to be
unitarily equivalent to the multiplication operator given by the
rational function $-p(i\kappa)^*/p(i\kappa)$, 
the incoming and 
outgoing  translation representations and the
Lax-Phillips semigroup, which describes the evolution of the states
which are neither incoming in the past nor outgoing in the future.
\end{abstract}

\maketitle

\begin{section}{Introduction}
In this paper we investigate the spectral theory, dynamics and Lax-Phillips
scattering for an abstract system which models the interaction between
a finite dimensional linear subsystem and an infinite
dimensional wave field on a halfline. We will call such systems {\it
generalized Lamb models} in that they extend the standard Lamb model
(see \cite{[L]}) that will be introduced shortly. 
Although our main concern is with linear oscillators, we will describe some properties of the models in the anharmonic case also.\par
To introduce our models, let us consider a $n$-dimensional lagrangian system linearized around a certain equilibrium point. Its equations of motion are given by
$$
G\ddot y=H y
$$
where $y\in \RE^n$ is the displacement from the given equilibrium point (for reference, $y=0$), 
and the matrices $G$ and $H$  represent the quadratic approximation of kinetic and potential 
energy around the equilibrium point. $G$ is positive definite and both matrices are symmetric 
with respect to the standard inner product in $\RE^n$. For technical and theoretical reasons it 
is more useful to endow $\RE^n$ with the inner product given by $G$. With respect to this inner product the matrix $L=G^{-1}H $ is symmetric and the lagrangian equation take the form

$$
\ddot y=L y
$$
with $L$ symmetric with respect to the $G$ inner product. 
The case of a chain of harmonic oscillators is well known and yields to a Jacobi matrix for the operator $L$.
\par
Analogously, let us consider the wave equation on the halfline. To be
definite let us consider Neumann boundary condition at the
origin. Denoting with $\Delta_N$ the 1-dimensional laplacian 
with homogeneous Neumann boundary conditions at the origin, the wave
field (we have posed equal to one the wave velocity) evolves according
to the wave equation

$$
 \ddot \phi=\Delta_N \phi \,.
$$
So we have two decoupled second order equations for two different oscillating systems, the finite dimensional one with generator $L$ and the infinite dimensional one with generator $\Delta_N$. \par
Thus on the direct sum $L^2(\RE_+)\oplus\RE^n$ we have the
self-adjoint operator $$A_0=\Delta_N \oplus L$$
and the corresponding abstract wave equation 
$$
\ddot \Psi=A_0\Psi \,.
$$
In a heuristic way a coupling between the two oscillating systems 
could be given by posing a constraint between boundary values of 
the wave field at the origin and the displacement of the finite 
dimensional system.
The prototype of this coupling is fournished by the well known Lamb
model where a semi-infinite string is coupled  to a single particle oscillating in the transverse
 direction (see section 4.1 for the general case of a chain of oscillators); 
the particle, with mass $M$, is subjected to the tension $T$ of the string at the origin and to a restoring harmonic force with spring constant $K$, so that the formal equations are given by the system
\begin{align*}
\ddot \phi(t,x)=&\phi''(t,x) \qquad x>0 \, , \\
M \ddot y(t)=& -Ky(t)+T\phi'(t,0_+)\, .
\end{align*}
plus the constraint $$\phi(t,0_+)=y(t)\,.$$ 
This model was proposed by Horace Lamb in 1900 
as an example of dissipation in (subsystems of) conservative systems. In fact it is possible to decouple field and particle dynamics, and the particle component satisfies a reduced equation which turns out to be, for $t>0$,
$$
M\ddot y(t)+2T\dot y(t)+ K y(t) = 
T(\phi'_0(t,0_+) + \dot \phi_0(t,0_+)) 
$$
The forcing term on the right hand side depends on the evaluation at
the origing of the {\it free} evolution for the wavefield 
of the initial data $\phi_0$, $\dot \phi_0$. Thus for initial data of compact
support the forcing term is a pure transient definitively vanishing,
and the reduced dynamics for the particle coincides for large times
with that of a damped harmonic oscillator, so that the effect of
interaction between particle and field reduces to damping only. This
means exponentially fast return to equilibrium of the finite
dimensional subsystem and correspondingly a neat transfer of energy to
the field. This relaxation property towards the equilibrion position
of the finite dimensional component is always true when 
the corresponding self-adjoint operator has
empty point spectrum (see Remark \ref{equi}) as it is the case in the Lamb model. 
The result has various generalizations to anharmonic oscillators (see \cite{[K]}). 
\par\noindent
Some tridimensional models reduce themselves to generalized Lamb models due to simmetry.
The case of an elastic spherical shell coupled to the acoustic field when radial oscillations 
only are allowed is treated in section 4.3, and it yelds to a nontrivial generalized Lamb model. 
Another issue of interest of these coupled systems are given by the
fact that some linear models of classical and quantum field theory
reduce themselves in the ultraviolet limit, and after due
renormalizations, to generalized Lamb models. For example, the
Schwabl-Thirring (see \cite{[ST]}) model when restricted to its monopole sector (the only one where it is not trivial) and after a spring constant renormalization turns out to be equivalent to a Lamb model (see \cite{[LM]}).
A similar phenomenon occurs for the Pauli-Fierz model describing the
interaction of a charged oscillator with the electromagnetic field in
dipole approximation and after mass renormalization (see \cite{[BNP]}) . In this case, however, reduction of the dynamics on its non trivial part, yields a boundary condition different from that of the Lamb model (see Example \ref{Pauli-Fierz}). \par
This discussion of motivating examples, and relevant studies existing in the literature, shows however that the usual formulation is partly formal in that it is not clear what it should be the functional setting of the Lamb system in the first place, and secondarily its Hamiltonian formulation, if any exists. \par
A guide to set rigorously these questions in this and in  more general
situations is suggested by an analysis of the coupling between field
and particle. The idea is to restrict the uncoupled vector operator
$A_0$ to the linear variety defined by the constraint existing between
field and subsystem; the uncoupled operator on this linear variety is
no more self-adjoint but it is symmetric with defect indices $(1,1)$. All
possible selfadjoint extensions different from $A_0$ itself correspond
to a well defined coupling or interaction between two subsystems.  The
case of the Lamb model, for example, corresponds  to the closed linear
variety $\phi(0_+)=y$. The most general boundary conditions still
producing a self-adjoint operator, as we
will see, is of the kind $\theta\phi'(0_+)+\phi(0_+)= w \cdot y$, 
where $\theta\in\RE$ and $w$ is a given vector in $\RE^n$. 
The case of a chain of harmonic oscillators one of which coincides
with the boundary point of the string, which is the more obvious
generalization of the Lamb model, corresponds to a vector $w$ with
just a single entry nonvanishing and to $\theta=0$. A generic $w$ corresponds to nonlocal coupling between string and more than one oscillator, i.e. the interaction is not "nearest neighbour". 
Our first concern, is to give a rigorous account of this construction
and to explicitely describe the interacting system so obtained (see 
Theorem \ref{estensioni}) as well as its spectral properties (see
Theorem \ref{spectrum}). The
interacting operator so costructed is a singular perturbation of the
selfadjoint operator $A_0$, related to the class of one dimensional
point interactions, or better point interaction with inner structure 
previously studied in different context and with a different formalism
by many authors 
(see e.g. \cite{[Pa]}, \cite {[Ku]}, \cite{[AK]} and references therein). 
In passing, we note that the coupled operator we study corresponds to
a boundary value problem for the wavefield only, but with an
eigenparameter dependent boundary condition (see Remark \ref{boundary}); this sort of parameter dependent boundary value problems are well known in the literature, both physical and mathematical. However we do not follow this road to the study of spectral and scattering properties of the coupled operator. \par
As a byproduct of the construction we obtain in Section 3 the
Hamiltonian structure of the system (see Theorem \ref{group}), 
which we generalize to
the case of anharmonic oscillators, giving conditions for the
existence of global flow (see Remark \ref{nonlinear}). 
As far as we know, a completely rigorous description of the
Hamiltonian structure of such type of systems has been lacking up to
now, whereas interesting, but formal treatments, are scattered in the
literature 
(see for example \cite {[LM]}, \cite {[HBW]}).\par
In the case the symmetric operator $L$ has no degenerate eigenvalues 
we show that the dynamics of the system is equivalent, for a dense
set of smooth initial data, to a reduced dynamics of a
wave equation on the half line which incorporates the interaction with
the finite dimensional systems through a higher order boundary
condition of the kind $p(\partial_x)\phi(t,0_+)=0$, where
the polynomial $p$ is explicitely related to the paramenters entering
into the definition of the model (see Theorem \ref{ponte}). 
This is a technical result, useful for the analysis of the Lax-Phillips scattering for the system, which is the main topic of the remaining part of the paper. \par 
In Section 6, in the case of empty point specrum, 
we determine the incoming ($R^-$) and outgoing ($R^+$) 
translation representations which make the dynamics unitarily equivalent to the
translation on $L^2(\RE)$ defined by $T^tf(x):=f(x-t)$. This provides the scattering operator $S^*_p$ for the
system by the relation $S^*_p=R^+(R^-)^{-1}$. Moreover 
$S^*_p$ turns out to be unitarily equivalent 
to the multiplication operator given by the rational function 
$-p(i\kappa)^*/p(i\kappa)$. 
In Section 7 the Lax-Phillips semigroup $Z^t$, $t\ge 0,$ which describes the evolution
of the states which are neither incoming in the past nor outgoing in
the future is completely characterized. It acts on a finite
dimensional vector space, 
whose dimension coincides with the degree of the polynomial $p$, 
by $Z^t=e^{-tB}$,
where the spectrum of the generator $B$ is made of the resonances of
the system. Such resonances correspond to the roots of the
polynomial $p$.
\end{section}

\begin{section}{Singular perturbations of the free dynamics}

Let us begin with some definitions. We 
denote by $L^2(\RE_+)$ the Hilbert space of square-integrable functions
on  the half-line $(0,+\infty)$ and by $H^1(\RE_+)$ and $H^2(\RE_+)$
the Sobolev spaces 
$$H^1(\RE_+):=\left\{\phi\in
L^2(\RE_+)\,:\, \phi'\in L^2(\RE_+)\right\}\,,$$
$$
H^2(\RE_+):=\left\{\phi\in
L^2(\RE_+)\,:\, \phi',\,\phi''\in L^2(\RE_+)\right\}\,.$$
Here the prime $\phi'$ denotes a spatial derivative. With a dot, $\dot\phi$,
we will denote a time derivative. 
We then define $H_N^2(\RE_+)$ as the subspace of $H^2(\RE_+)$ of
functions which satisfy homogeneous Neumann boundary conditions at
zero, i.e. $$H_N^2(\RE_+):=\left\{\phi\in
H^2(\RE_+)\,:\, \phi'(0_+)=0\right\}\,.$$
We denote by $\langle\cdot,\cdot\rangle_2$ and by $\|\cdot\|_2$ the usual
scalar product and the corresponding norm on $L^2(\RE_+)$.\par
Given the $n$-dimensional Hilbert space $\fh$ with scalar product
$\langle\cdot,\cdot\rangle$ and corresponding norm $\|\cdot\|$, and given 
the symmetric operator $L:\fh\to\fh$, 
we consider the self-adjoint operator 
$$A_0:H^2_N(\RE_+)\oplus\fh\subset L^2(\RE_+)\oplus\fh\to L^2(\RE_+)
\oplus\fh\,,\quad A_0(\phi,y):=(\phi'',L y)\,.$$
Regarding the spectrum of $A_0$ one has 
$$
\sigma_{ess}(A_0)=\sigma_{ac}(A_0)=(-\infty,0]\,,\quad 
\sigma_{pp}(A_0)=\sigma(L)\,.
$$
In order to couple the two dynanical sistems described by the
equations $\ddot\phi=\phi''$ and $\ddot y=Ly$ we define the
continuous and surjective linear operator
$$
\tau:H^1(\RE_+)\oplus\fh\to\CO\,,\quad 
\tau(\phi,y):=\phi(0_+)-\langle w,y\rangle\,,\qquad w\in\fh\,,
$$
and then we consider the closed symmetric operator $\dot A_0$ obtained by
restricting $A_0$ to the kernel of $\tau$. $\dot A_0$ has deficiency
indices $(1,1)$ and we are interested in its self-adjoint
extensions different from $A_0$ itself, which we parametrize by the real
extension parameter $\theta$. Thus to each quadruple 
$(\fh, L, w,\theta)$ corresponds a
different {\it generalized Lamb model}. The next theorem completely
characterizes such models. 
 
\begin{theorem}\label{estensioni} For any $\theta\in\RE$ 
the linear operator 
$$A:D(A)\subset L^2(\RE_+)\oplus\fh\to L^2(\RE_+)\oplus\fh $$ defined by 
$$
D(A):=\left\{(\phi,y)\in H^2(\RE_+)\oplus\fh\,:\, \theta\phi'(0_+)
+\phi(0_+)=\langle w,y\rangle\right\}\,,
$$
$$
A(\phi,y):=(\phi'', L y+ w\, \phi'(0_+))
$$
is a self-adjoint extension of $\dot A_0 $ and its resolvent is given by
$$
(-A+z)^{-1}=(-A_0+z)^{-1}+(\theta+\Gamma(z))^{-1}G_z\otimes
G_{z^*}\,,
$$
where 
$$
\Gamma(z):=-\left(\pm\,\frac{1}{\sqrt z}+\langle w,(-L+z)^{-1}w\rangle\right)\,,
\quad \pm\text{\rm Re}\sqrt z>0
$$
and
$$
G_z=\left(\pm\frac{e^{\mp\sqrt z\,x}}{\sqrt z},\,- (-L+z)^{-1}w
\right)\,,\quad \pm\text{\rm Re}\sqrt z>0
\,.
$$
\end{theorem}
\begin{proof} We will make use of the mathematical procedure developed
in \cite{[P]} (see also \cite{[CFP]}, Theorem 2.2, for a similar proof in the case
of a one-dimensional model in acoustics).\par
For any $z\in\rho(A_0)$, let us consider the two linear
continuous operators
$$
\breve G(z):L^2(\RE_+)\oplus\fh\to\CO\,,\quad\breve G(z):=\tau(-A_0+z)^{-1}\,,
$$
$$
G(z):\CO\to L^2(\RE_+)\oplus\fh\,,\quad G(z):= \breve G(z^*)^*\,.
$$
Since $$
\left(-\frac{d^2}{dx^2}+z\right)^{-1}:L^2(\RE_+)\to H_N^2(\RE_+)$$ has kernel 
$$
{\mathcal G}_N(z;x_1,x_2)=\pm\frac{e^{\mp\sqrt z\,|x_1-x_2|}+e^{\mp\sqrt
z\,(x_1+x_2)}}{2\sqrt z}\,,\quad \pm\text{\rm Re}\sqrt z>0\,,
$$ 
the operators $\breve G(z)$ and $G(z)$ are 
represented by the vectors $G_{z^*}$ and $G_z$ respectivelty,
where
$$
G_z=\left(\pm\frac{e^{\mp\sqrt z\,x}}{\sqrt z},\,- (-L+z)^{-1}w
\right)\,,\quad \pm\text{\rm Re}\sqrt z>0
\,.
$$
Note that 
\begin{equation}\label{range}
\text{\rm Ran}(G(z))\cap D(A_0)=\left\{0\right\}\,.
\end{equation}
Now we define, 
for any $z\in\rho(A_0)$, the map
$$
\Gamma(z):\rho(A_0)\to\CO\,,\quad \Gamma(z):=-\tau G(z)\,,
$$ 
i.e.
$$
\Gamma(z):=-\left(\pm\,\frac{1}{\sqrt z}+\langle w,(-L+z)^{-1}w\rangle\right)\,,
\quad \pm\text{\rm Re}\sqrt z>0\,,\,.
$$
At first let us note that the
the function $\Gamma$ satisfies the relation 
\begin{equation}\label{gamma}
\Gamma(z)-\Gamma(w)=(z-w)\,\breve G(w)G(z)\,.
\end{equation}
Indeed 
$$
\Gamma(z)-\Gamma(w)=\tau(G(w)-G(z))
$$
and, by first resolvent identity and by the definition of $G(z)$,
$$
(z-w)\,(-A_0+z)^{-1}G(z)=G(w)-G(z)\,.
$$
Relation (\ref{gamma}) implies that 
$$
R(z):=(-A_0+z)^{-1}+(\theta+\Gamma(z))^{-1}G_z\otimes
G_{z^*}
$$
satisfies the first resolvent equation
\begin{equation}\label{resolvent}
R(w)- R(z)=(z-w)\,R(w) R(z)\,.
\end{equation}
By the definitions of $\breve G(z)$ and $G(z)$, and since
$\Gamma(z)^*=\Gamma(z^*)$, one obtains
\begin{equation}\label{symm}
R(z)^*=R(z^*)\,.
\end{equation}
Moreover, by (\ref{range}), the linear
operator $R(z)$ is injective. Thus 
$$
A:=-R(z)^{-1}+z
$$
is well defined on the domain
$$
D(A):=\text{\rm Range$(R(z))$}\,.
$$
By (\ref{resolvent}) such a definition is $z$-independent. By (\ref{symm}) $A$ is
symmetric and is self-adjoint since 
$$
\text{\rm Range$(-A\pm i)=L^2(\RE_+)\oplus\fh$}
$$
by construction. We have thus defined the self-adjoint operator
$$
D(A):=\left\{(\phi_z,y_z)+(\theta+\Gamma(z))^{-1}(\phi_z(0_+)
-\langle w, y_z\rangle)\,G_z\,,\ \phi_z\in H_N^2(\RE_+)\right\}\,,
$$
$$
(-A+z)(\phi,y):=(-A_0+z)(\phi_z,y_z)\,.
$$
This implies 
$$
\phi'(0_+)=-(\theta+\Gamma(z))^{-1}(\phi_z(0_+)
-\langle w, y_z\rangle)
$$
and 
$$
\phi(0_+)=\phi_z(0_+)\mp\frac{1}{\sqrt z} \,\phi'(0_+)\,.
$$
Therefore
\begin{align*}
&\theta \phi'(0_+)=(\theta+\Gamma(z))\,\phi'(0_+)-\Gamma(z)\,\phi'(0_+)\\
=&-\phi_z(0_+)+\langle w, y_z\rangle
+\left(\frac{\pm 1}{\sqrt z} +\langle w,(-L+z)^{-1}w\rangle\right)\,\phi'(0_+)\\
=&-\phi'(0_+)+\langle w,\left(y_z+(-L+z)^{-1}w\rangle\,\phi'(0_+)\right)\\
=&-\phi(0_+)+\langle w,y\rangle\,.
\end{align*}
Posing
$$
A(\phi,y)\equiv\left(A_1(\phi,y),A_2(\phi,y)\right)
$$
one obtains
\begin{align*}
&A_1(\phi,y)(x)=\phi''_z(x)\mp z\phi'(0_+)\,\frac{e^{\mp\sqrt
z\,x}}{\sqrt z}\\
=&\left(\phi_z(x)\mp\phi'(0_+)\,\frac{e^{\mp\sqrt
z\,x}}{\sqrt z}\right)''=\phi''(x)
\end{align*}
and
\begin{align*}
&A_2(\phi,y)=L y_z+z\phi'(0_+)\,(-L+z)^{-1} w\\
=&L y+\left(-L(-L+z)^{-1}+z(-L+z)^{-1}\right)w\, \phi'(0_+)\\
=& L y+w\, \phi'(0_+)
\,.
\end{align*}

\end{proof}
Let us define the (eventually
empty) set 
$$
\sigma_w(L):=\{\lambda\in\sigma(L)\,:\,w\in\fh_\lambda^\perp\}
\,,$$
where $\fh_\lambda$ denoted the spectral subspace relative to $\lambda$.
\begin{theorem} \label{spectrum} 
$$\sigma_{ess}( A)=\sigma_{ac}(A)=(-\infty,0]\,,$$
$$
\sigma_{pp}(A)=\sigma_w(L)\cup\left\{\lambda\in\rho(L)\cap (0,+\infty )\,:\,
\theta+\Gamma(\lambda)=0\right\}\,.$$
\end{theorem}
\begin{proof}
The properties regarding the essential and continuous spectrum are
more or less standard and can be obtained proceding as in
\cite{[CFP]}, Theorem 2.3. Let us now come to the point spectrum.
\par\noindent
1. Let $\lambda\in\sigma(L)$. Then $(0,y_\lambda)$ 
is an eigenvector if $y_\lambda$ solves the equations
$$
L y_\lambda=\lambda y_\lambda\,,\qquad\langle w,y_\lambda\rangle=0\,,
$$
thus $\lambda\in\sigma_w(L)$.\p
2. 
Let $\lambda>0$. Then $\phi_\lambda(x):=e^{-\sqrt \lambda\,x}$ solves
$\phi_\lambda''=\lambda\phi_\lambda$. Thus 
$(\phi_\lambda,y_\lambda)$ is an eigenvector if $\lambda$ and $y_\lambda$ solve the 
equations 
\begin{equation}\label{eigenv}
L y_\lambda-\sqrt\lambda\,w=\lambda y_\lambda\,,
\end{equation}
$$
-\theta\sqrt\lambda+1-\langle w, y_\lambda\rangle=0\,.
$$
If $\lambda\in\rho(L)$ then 
$$y_\lambda=-\sqrt\lambda\,(-L+\lambda)^{-1}w$$
and $\lambda$ must solve the equation
$$-\theta\sqrt\lambda+1+\sqrt\lambda\,\langle w,(-L+\lambda)^{-1}w\rangle=0\,.
$$
If otherwise $\lambda\in\sigma(L)$ then (\ref{eigenv}) can be solved only if
$w\in\fh_\lambda^\perp$ by 
$y_\lambda=y_\lambda^\parallel+y_\lambda^\perp$, where
$y_\lambda^\parallel\in\fh_\lambda$ and  $y_\lambda^\perp\in\fh_\lambda^\perp$
is defined by
$$
y_\lambda^\perp:=-\sqrt\lambda\,(-L_\lambda+\lambda)^{-1}w\,,\quad
L_\lambda:=(1-P_\lambda) L
(1-P_\lambda):\fh_\lambda^\perp\to\fh_\lambda^\perp\,.
$$
Thus $\lambda\in\sigma_w(L)$ and moreover it has to solve the
equation
$$
-\theta\sqrt\lambda+1+\sqrt\lambda\,\langle
w,(-L_\lambda+\lambda)^{-1}w\rangle=0\,.
$$
\end{proof}
\begin{remark}
When $\sigma_w(L)$ is empty, i.e. in the generic situation, the point
  spectrum of the interacting operator $A$ is quite different from the
  point spectrum of the decoupled one, $A_0$. In particular, the free
  eigenvalues of the finite dimensional subsystem disappear, and in their place could possibly appear the real solutions of the equation $\Gamma(\lambda)+\theta=0$. In fact, as we shall see in Section 6, the disappeared eigenvalues, which for the uninteracting operator $A_0$ are immersed in the continuum spectrum, become resonances of the interacting operator. 
\end{remark}

\begin{remark}
In the case $\sigma_w(L)\not=\emptyset$ we can suppose, 
without loss of generality,
  $\fh=\fh_1\oplus\fh_2$, $L=L_1\oplus L_2$ and $w=w_1\oplus 0$.
Then for the self-adjoint extensions given in Theorem
\ref{estensioni} we have $A=A_1\oplus L_2$,
where
$$A_1:D(A_1)\subset L^2(\RE_+)\oplus\fh_1\to L^2(\RE_+)\oplus\fh_1 $$ is 
defined by 
$$
D(A_1):=\left\{(\phi,y_1)\in H^2(\RE_+)\oplus\fh_1\,:\, \theta\phi'(0_+)
+\phi(0_+)=\langle w_1,y_1\rangle_{\fh_1}\right\}\,,
$$
$$
A_1(\phi,y_1):=(\phi'', L_1y_1+ w_1\, \phi'(0_+))\,,
$$
i.e. the dynamics on $\fh_2$ is trivial, in the sense that it is
decoupled from the field one. Thus the hypothesis
$\sigma_w(L)\not=\emptyset$  is equivalent to the existence of a
subspace on which the particles and the field are uncoupled. In other words 
the points of the pure point spectrum belonging to 
$\sigma_w(L)$ correspond to "radiationless
motions", in which the interaction between oscillators and field is
decoupled. Similar exceptional solutions in which the subsystem
oscillates at a normal frequency of the decoupled system are known, 
for example, also in the classical electrodynamics of an extended
charge where they are called Bohm-Weinstein modes. In that case the
coupling between field and particle is defined by the charge density $\rho(x)$ of the particle, and the condition to have radiationless modes of frequency $\omega$ is that the Fourier transform of the form factor satisfies $\hat{\rho}(\omega)=0$. 
\end{remark}
\vskip10pt
\begin{remark}\label{boundary}
The operator $A$ can be interpreted in a formal way as a differential operator with an eigenvalue dependent boundary condition. Let us consider the secular equation for the operator $A$ and in particular its finite dimensional component, and couple it with the boundary condition for elements of the domain of the operator. We get 
\begin{align*}
L y+ w\, \phi'(0_+)=& \lambda y\\
\theta\phi'(0_+)+\phi(0_+)=&\langle w,y\rangle
\end{align*}
From the first equation it follows $y=-\phi'(0_+)(L-\lambda)^{-1} w$ and substituting in the second equation one gets
$$
\left(\theta+\left\langle w,(L-\lambda)^{-1} w\right\rangle\right)\phi'(0_+) + \phi(0_+) = 0
$$
which is, formally, a Robin boundary condition for the field at the origin. The condition contains the eigenvalue $\lambda$ and it is is known in the physical and mathematical literature as an energy dependent boundary condition.
From this point of view, the boundary value problem for the coupled operator can be reduced to a boundary value problem for the field only, but eigenvalue dependent.
\par
By the way note that, as it should be, the above eigenvalue dependent boundary condition is equivalent to the eigenvalue equation in Theorem 2.3, $\Gamma(\lambda)+\theta=0$, as it is immediately seen by the position $\sqrt z = \lambda$. 
\end{remark}
\end{section}

\begin{section}{The Hamiltonian structure}
In this section we are interested in describing the Hamiltonian
structure of the dynamical system related to the abstract wave
equation
$$\ddot\Psi=A\Psi\,,\qquad \Psi\equiv(\phi,y)\,.$$
The solution of the corresponding Cauchy problem is then given by the
symplectic flow generated by the linear operator 
$\begin{pmatrix} 
0&1\\
A&0 
\end{pmatrix}$.\par
We begin by determinig the quadratic form corresponding to $-A$:
\begin{theorem} Let us denote by $\Q$ the quadratic form of
$-A$. \par\noindent 
1. If $\theta=0$ then
$$
D(\Q)=\left\{(\phi,y)\in H^1(\RE_+)\oplus\fh\,:\,\phi(0_+)=\langle 
w,y\rangle\right\}\,,$$

$$ \Q: D(\Q)\to\RE\,,\quad\Q(\phi,y)=\|\phi'\|^2_2-\langle
L y, y\rangle\,.
$$
2. If $\theta\not=0$ then $D(\Q)=H^1(\RE_+)\oplus\fh$ and
$$
\Q: H^1(\RE_+)\oplus\fh\to\RE\,,\quad
\Q(\phi,y)=\|\phi'\|^2_2-\langle L y, y\rangle
-\frac{1}{\theta}\,\left|\phi(0_+)-\langle w,y\rangle\right|^2\,.
$$
\end{theorem}
\begin{proof} For any $(\phi,y)\in D(A)$ one has
\begin{align*}
\Q(\phi,y)=&\|\phi'\|^2_2-\langle L y,
y\rangle+(\phi')^*(0_+)(\phi(0)-
\langle w, y\rangle)\\
=&\|\phi'\|^2_2-\langle L y,y\rangle-\theta|\phi'(0_+)|^2\,.
\end{align*} 
Thus the proof is done if $\Q$ is bounded from below and
closed. This follows from
$$
|\phi(0_+)|^2\le a\,\|\phi\|^2_2+b\,\|\phi'\|^2_2\,,\quad a>0,\ 0<b<1\,.
$$ 
\end{proof}
Let us make $D(\Q)\subset L^2(\RE^3)\oplus\fh$ a Banach
space with norm 
$$\|(\phi,y)\|_Q^2
:=\Q(\phi,y)+(\sup\sigma(A)+1)(\|\phi\|^2_2+\|y\|^2)$$ and define 
$$
\H_\circ:=D(Q)\oplus L^2(\RE_+)\oplus \fh\,.
$$
Then one has the following 
\begin{theorem} \label{group} The linear operator 
$$
\begin{pmatrix} 
0&1\\
A&0 
\end{pmatrix}
:D(A)\oplus D(\Q)\subset\H_\circ\to \H_\circ\,,
$$
is the generator of a strongly continuous group of evolution $$
U_\circ^t:\H_\circ\to\H_\circ$$ which preserves the energy 
$$\E((\phi,y),(\dot\phi,\dot y)):=\frac{1}{2}\,
\left(\Q(\phi,y)+\|\dot\phi\|^2_2+\|\dot y\|^2\right)\,.$$
Such an operator is the Hamiltonian linear vector field corresponding to the quadratic Hamiltonian $\E$
via the canonical symplectic form on
$L^2(\RE_+)\oplus\fh\oplus L^2(\RE_+)\oplus\fh$ given by
\begin{align*}
&\Omega((\phi_1,y_1,\dot\phi_1,\dot y_1),(\phi_2,y_2,\dot\phi_2,\dot
y_2))\\
:=&\langle\phi_1,\dot\phi_2\rangle_2-\langle\phi_2,\dot\phi_2\rangle_2
+\langle y_1,\dot y_2\rangle-\langle y_2,\dot y_2\rangle\,.
\end{align*}
\end{theorem} 
\begin{proof} The operator $A$ is self-adjoint and bounded from
above. Thus the result concerning evolution generation follows from the theory of abstract
wave equations (see e.g. \cite{[G]}, chapter 2, section 7). The results
about the Hamiltonian structure follows from the theory of linear
Hamiltonian systems in infinite dimensions (see e.g. \cite{[CM]},
chapter 2).
\end{proof}

\begin{remark}\label{nonlinear} The above results can be immediately extended to a nonlinear
situation. Indeed, given the potential function $V$ let
us consider the Hamiltonian
$$H:\H_\circ\to\RE\,,$$ 
where
$$ 
H((\phi,y),(\dot\phi,\dot y))
=\frac{1}{2}\,\left(\|\phi'\|^2_2+\|\dot\phi\|_2^2+\|\dot
y\|^2\right)+V(y)
$$
when $\theta=0$ and
$$
H((\phi,y),(\dot\phi,\dot y))
=\frac{1}{2}\,\left(\|\phi'\|^2_2+\|\dot\phi\|_2^2+\|\dot
y\|^2-\frac{1}{\theta}\,\left|\phi(0_+)-\langle
w,y\rangle\right|^2\right)+
V(y)
$$
when $\theta\not= 0$. \par
The non linear Hamiltonian vector field corresponding,
via the canonical symplectic form on
$L^2(\RE_+)\oplus\fh\oplus L^2(\RE_+)\oplus\fh$, to $H$ is given by
$$X_H:D(A)\oplus D(Q)\subset \H_\circ
\to \H_\circ\,,$$ 
$$
X_H((\phi,y),(\dot\phi,\dot y)):=(\dot\phi,\dot y,
\phi'', -\nabla V(y)+ w\, \phi'(0_+))\,.
$$
Obviously $X_H=X_E+B$, where $X_E$ is the linear Hamiltonian vector
field corresponding to the quadratic Hamiltonian $E$ and $B$ is vector
field $B((\phi,y),(\dot\phi,\dot y)):=(0,0,-(Ly+\nabla V(y)),0)$. Thus if
$V$ is twice differentiable then by Segal's existence theorem (see
\cite{[S]}) $X_H$ generates a local continuous non linear symplectic flow on 
$\H_\circ$. Since $Q$ is bounded from below, if 
$$
V(y)\ge c_1\|y\|^2-c_2\,,\qquad
c_1>0\,,\quad c_2\ge 0\,,$$ then such a flow is global. 
\end{remark}
\end{section}

\begin{section}{Examples} 
\begin{example} \label{lamb} The dynamics of the Lamb model (see
\cite{[L]}) we already described in the introduction, given by the equations 
\begin{align*}
\ddot\phi(t,x)&=\phi''(t,x)\\
M\ddot y(t)&=-Ky(t)+T\phi'(t,0_+)\\
y(t)&=\phi(t,0_+)
\end{align*}
is described by the self-adjoint
extension $A$ corresponding to
$$\text{\rm dim}\,\fh=1\,,\quad 
\langle x,y\rangle=\frac{M}{T}\,x^*y\,,
\quad Ly=-\frac{K}{M}\,y\,,\quad
w=\frac{T}{M}\,,\quad \theta=0 
\,.
$$ 
The similar model with $n$ point masses
\begin{align*}
\ddot\phi(t,x)&=\phi''(t,x)\\
M_1\ddot y_1(t)&=-K_1(y_1(t)-y_{2}(t))+T\phi'(t,0_+)\\
M_{2}\ddot y(t)_{2}&=-K_{2}(y_{2}(t)-y_{3}(t))
+K_1(y_1(t)-y_{2}(t))\\
\vdots&\\
M_{n-1}\ddot y_{n-1}(t)&=-K_{n-1}(y_{n-1}(t)-y_n(t))
+K_{n-2}(y_{n-2}(t)-y_{n-1}(t))\\
M_n\ddot y_n(t)&=-K_ny_n(t)+K_{n-1}(y_{n-1}(t)-y_{n}(t))\\
y_1(t)&=\phi(t,0_+)
\end{align*}
is described by the self-adjoint
extension $A$ corresponding to
$$
\text{\rm dim}\,\fh=n\,, \quad 
\langle x,y\rangle=\frac{1}{T}\sum_{j=1}^{n}M_j\,x^*_jy_j\,, 
$$ 
$$\label{matricelamb} L=
\begin{pmatrix} 
-\frac{K_1}{M_1}&\frac{K_1}{M_1}&0&0&\hdots &0\\
\frac{K_1}{M_2}&-\frac{K_1+K_2}{M_2}&\frac{K_2}{M_2}&0&\hdots &0\\
0&\frac{K_2}{M_3}&-\frac{K_2+K_3}{M_3}&\frac{K_3}{M_3}&\hdots &0\\
\vdots&\ &\ &\ddots &\ &\vdots\\
0&\hdots &\frac{K_{n-3}}{M_{n-2}}&-\frac{K_{n-3}+K_{n-2}}{M_{n-2}}
&\frac{K_{n-2}}{M_{n-2}}&0\\
0&\hdots &0&\frac{K_{n-2}}{M_{n-1}}&-\frac{K_{n-2}+K_{n-1}}{M_{n-1}}
&\frac{K_{n-1}}{M_{n-1}}\\
0&\hdots &0&0&\frac{K_{n-1}}{M_{n}}&-\frac{K_{n-1}+K_n}{M_{n}}\\
\end{pmatrix}\,,
$$
$$
w=\left(\frac{T}{M_1},0,\dots,0\right)\,,\quad\theta=0\,.
$$
The corresponding Hamiltonian is given by
$$H:\left\{(\phi,y)\in
H^1(\RE_+)\oplus\CO^n\,:\,\phi(0_+)=y_1\right\}\oplus
L^2(\RE_+)\oplus\CO^n\to\RE$$
$$
H((\phi,y),(\dot\phi,\dot y)):=
\frac{1}{2}\left(\|\dot\phi\|^2_2 + \|\phi'\|^2_2 + \frac{1}{T}\,\sum_{k=1}^{n}M_k|\dot y_k|^2
+\frac{1}{T}\,\Lambda y\! \cdot\! y\right)\,,
$$
where the matrix $\Lambda$ is given by 
$$ \Lambda=
\begin{pmatrix} 
{K_1}&{-K_1}&0&\hdots &0\\
{-K_1}&{K_1+K_2}&{-K_2}&\hdots &0\\
\vdots&\ &\ddots &\ &\vdots\\
0&\hdots &{-K_{n-2}}&{K_{n-2}+K_{n-1}}
&{-K_{n-1}}\\
0&\hdots &0&{-K_{n-1}}&{K_{n-1}+K_n}\\
\end{pmatrix}
$$
and $\cdot $ denotes the standard inner product in $\CO^n$.

\end{example}
In the following examples we describe models, from classical
electrodynamics and theoretical acoustic respectively, which are not
interpretable as standard Lamb models in that $\theta\not=0$.
\begin{example}\label{Pauli-Fierz} The renormalized Pauli-Fierz model.
\par
A three dimensional charged oscillator characterized by frequency $\omega$, mass
$m$ and electric charge $\text{e}$ interacting with the
electromagnetic field in dipole approximation has a dynamics
described, in the point limit and after mass renormalization, by a
well defined self-adjoint operator which couples 
particle momentum and vector electromagnetic potential. Dynamics and its main properties, classical and quantum, are constructed and studied in \cite{[BNP]}. This is the point limit of the Pauli-Fierz model for the case of a quadratic potential energy. 
Due to the dipole approximation, the action of this operator is non
trivial (i.e. different from the free uncoupled dynamics) only on the 
radial component of the field, and by standard decomposition using vector spherical harmonics it turns out that on this monopole subspace, the restricted dynamics for every couple $(\phi, p)$ constituted by a component of the vector potential on the non trivial subspace, and a corresponding component of the particle momentum, is given by the coupled system 
\begin{align*}
&\ddot\phi(t,r)=\phi''(t,r)\\
&\ddot p(t)=-\frac{3m}{2\text{e}}\,\omega^2\phi(t,0_+)\\
&\phi'(t,0_+)+ \frac{3m}{2{\text{e}}^2}\,
\phi(t,0_+)=\frac{1}{\text{e}}\, p(t)\,.
\end{align*}
Writing the field $\phi$ which appears in the evolution equation for $p$ in
terms of its derivative $\phi'$ and $p$ by means of the boundary
condition one obtains
\begin{align*}
&\ddot\phi(t,r)=\phi''(t,r)\\
&\ddot p(t)=-\omega^2 p(t) + {\text{e}}\,\omega^2\phi'(t,0_+)\\
&\frac{2\text{e}^2}{3m}\,\phi'(t,0_+)+ \phi(t,0_+)=\frac{2\text{e}}{3m}\, p(t)\,.
\end{align*}
Let us remark here that by Newton's law $\dot p=-m\omega^2 q$, where $q$
is a component of the particle position. 
Thus the Cauchy initial datum for $\dot p$
is obtained from the initial position.\par 
In conclusion the dynamics of the renormalized Pauli-Fierz model in
dipole approximation and with quadratic external potential 
is described by the self-adjoint operator $A$ corresponding to 
$$\text{\rm dim}\,\fh=1\,,\quad 
\langle x,y\rangle=\frac{2x^*y}{3m\omega^2}\,,
\quad Ly=-\omega^2\,y\,,\quad
w={\text{e}}\,\omega^2, \quad \theta=\frac{2\text{e}^2}{3m} 
\,.
$$ 
The corresponding Hamiltonian is given by
$$H:H^1(\RE_+)\oplus\CO\oplus
L^2(\RE_+)\oplus\CO\to\RE$$
\begin{align*}
&H((\phi,p),(\dot\phi,\dot p)):=
\frac{1}{2}\left(\|\dot\phi\|^2_2 + \|\phi'\|^2_2\right)\\ 
&+ \frac{1}{3m}\,\left(\frac{|\dot p|^2}{\omega^2}
+ |p|^2-
\left|\frac{3m}{2\text{e}}\,\phi(0_+)
-p\right|^2\right)\,.
\end{align*}
As recalled in the introduction, a field-particle 
interaction which reduces to the standard ($\theta=0$) Lamb model in
the point limit and after {\it spring constant} renormalization is the
Schwabl-Thirring model, in which a scalar field interacts with a
scalar oscillator 
(for details, in a different framework, see  \cite{[LM]}).
\end{example}
\begin{example}\label{acoustic} A spherical elastic shell in the acoustic field.\par
Let us consider the exterior problem for a spherical shell of mass
$M$, radius $R_0$
and constant surface density $\rho=\frac{M}{4\pi R_0^2}$ 
undergoing radial motion only, and
interacting with an irrotational acoustic field in the linear
approximation. The shell is elastic, i.e. on every surface element
acts a restoring force proportional to the radius, and of Young
modulus $K$. If small radial oscillations around $R_0$
are considered, introducing the variables $\psi$, related to the 
acoustic potential $\phi$ by $\phi(R_0+r) 
= \frac{\psi(r)}{r}$, $r>0,$ and the radius $R_0+R(t)$ and taking into
account the continuity of velocity $\phi'$ of the acoustic
field at the boundary one obtains the equations of motion (assuming 
propagation velocity equal to one) 
\begin{align*}
&\ddot\psi(t,r)=\psi''(t,r)\\
&M\ddot R(t)=-K R(t) + 4 \pi R_0 \rho\, \dot\psi(t,0_+)\\
&\frac{\psi'(t,0_+)}{R_0}-\frac{\psi(t,0_+)}{R_0^2}=\dot R(t)\,.
\end{align*}
This system can be converted to a generalized Lamb system by 
introducing the new variable (in fact a sort of total momentum)
$$
P:=M\dot R-4 \pi R_0 \rho\,\psi(0_+)\,.
$$
Re-writing the above equations in terms of the new variable $P$, 
expressing the field $\psi$ which appears in the evolution equation for $P$ in
terms of its derivative $\psi'$ and $P$ by means of the boundary
condition, one
obtains
\begin{align*}
&\ddot\psi(t,r)=\psi''(t,r)\\
&\ddot P(t)=-\frac{K}{M+4\pi R^3_0\rho}\,P(t)
-\frac{ 4 \pi  K R_0^2\rho}{M+4\pi R^3_0\rho}\,\psi'(t,0_+)\\
&-\frac{MR_0}{M+4\pi R^3_0\rho}\,\psi'(t,0_+)
+\psi(t,0_+)= -\frac{R_0^2}{M+4\pi R^3_0\rho}\,P(t)\,.
\end{align*}
As regards the initial datum for $\dot P$ it can be recovered from the
one for $R$ as in the
previous example.\par
By defining $\omega^2:=\frac{K}{M}$  
the above system is described by the
self-adjoint operator $A$ corresponding to 
$$\text{\rm dim}\,\fh=1\,,\quad 
\langle x,y\rangle=\frac{x^*y}{4\pi K}\,,
\quad Ly=-\frac{\omega^2}{1+R_0}\ y\,,$$
$$
w=-\frac{4\pi \omega^2 R^2_0}{1+R_0}\,,\quad \theta=-\frac{R_0}{1+R_0} 
\,.
$$ 
The corresponding Hamiltonian is given by
$$H:H^1(\RE_+)\oplus\CO\oplus
L^2(\RE_+)\oplus\CO\to\RE$$
\begin{align*}
&H((\psi,P),(\dot\psi,\dot P)):=
\frac{1}{2}\left(\|\dot\psi\|^2_2+ \|\psi'\|^2_2\right)\\
 &+ 
\frac{1}{8\pi M}\,\left(\frac{|\dot P|^2}{\omega^2}
+ \frac{|P|^2}{1+R_0}\right)+\frac{1+R_0}{2R_0}\,
\left|\psi(0_+)+\frac{R^2_0}{1+R_0}\,
\frac{P}{M}\right|^2\,.
\end{align*}
An analysis of interaction of elastic surfaces with acoustic fields 
from the point of view of Lax-Phillips scattering theory 
is given in \cite{[BE]}. For a study of one dimensional models in
acoustics in the framework of self-adjoint extensions we refer to \cite{[CFP]}.
\end{example}
\end{section}

\begin{section}{Wave equations with high-order boundary conditions}
From
now on we will consider self-adjoint operators $A$ which are
self-adjoint extensions corresponding (according to Theorem 
\ref{estensioni}) to $L$'s and $w$'s such that 
\begin{equation}\label{base}
\{L^{k}w\}_0^{n-1}\quad\text{\rm is a basis in
$\fh$.}\end{equation}
Note that the examples given in Section 4 satisfy such hypothesis.
\par
With respect to the orthonormal base obtained from
$\{L^kw\}_0^{n-1}$ by the Schmidt orthogonalization
procedure the linear operator $L$ is represented by a Jacobi
matrix. However we prefer to consider here the unitary
isomorphism $\fh\simeq\CO^n$ induced by the orthonormal
system $\left\{\hat e_i\right\}_{1}^n$ made of the
eigenvectors of $L$. For any vector $y\in\fh$ and
for any linear operator
$M:\fh\to\fh$ we pose 
$$y\equiv(y_1,\dots,y_n)\,,\qquad M\equiv\begin{pmatrix} 
M_{11}&\hdots &M_{1n}\\
\vdots&\hdots&\vdots\\
M_{n1}&\hdots &M_{nn}\,
\end{pmatrix}\,.
$$ 
With these notations 
$$
w\equiv(w_1,\dots,w_n)\,,\qquad
L\equiv
\begin{pmatrix} 
\lambda_1&0&\hdots &0\\
0&\lambda_2&\hdots &0\\
\vdots&\ &\ddots&\vdots\\
0&\hdots &0&\lambda_n\,
\end{pmatrix}\,,
$$
where $\sigma(L)=\left\{\lambda_1,\dots,\lambda_n\right\}$.
We introduce the diagonal matrix
$$
W\equiv
\begin{pmatrix} 
w_1&0&\hdots &0\\
0&w_2&\hdots &0\\
\vdots&\ &\ddots&\vdots\\
0&\hdots &0&w_n\,
\end{pmatrix}\,,
$$
the Vandermonde matrix
$$
V\equiv
\begin{pmatrix} 
1&1&\hdots &1\\
\lambda_1&\lambda_2&\hdots &\lambda_n\\
\vdots&\vdots &\hdots&\vdots\\
\lambda_1^{n-1}&\lambda_2^{n-1}&\dots&\lambda_n^{n-1}
\,
\end{pmatrix}
$$
and then we define 
$$M:=VW\,.$$
Since
$$ 
\det M=
\prod_{1\le i<j\le n}(\lambda_j-\lambda_i)\prod_{1\le i\le n}w_i\not=0\,,
$$ 
our hypothesis (\ref{base}) is equivalent to 
$$
\lambda_i\not=\lambda_j\quad\text{\rm and} \quad \sigma_w(L)=\emptyset.
$$
Thus under our hypothesis the spectrum of $L$ is simple and, 
by Theorem \ref{spectrum}, $A$ has no eigenvalue immersed
in the continuos spectrum.
\par
Let us denote by $\S(\RE_+)$ the space of rapidly decreasing smooth functions
on $[0,+\infty)$. We define the dense subspace $\A\subset\H_\circ$ by
\begin{align*}
\A:=\{(&\phi, y,\dot \phi,\dot y)\in\H_\circ\,:\,
\phi\in \S(\RE_+),\,\dot\phi\in \S(\RE_+),\\
&y=M^{-1}v(\phi),\, \dot y=M^{-1}v(\dot\phi)\}\,,
\end{align*}
where
$$
v(\phi)=\sum_{k=1}^n p_k(\partial_x)\phi\,(0_+)\,\hat e_k\,,
$$
and $p_k(\partial_x)$ is the differential operator with constant
coefficients associated to the polynomial recursively defined by
$$
p_1(z)=\theta z+1\,,\qquad
  p_{k}(z)=z^2p_{k-1}(z)-\langle w,L^{k-2}w\rangle z
\,,\quad k\ge 2\,.
$$
The next theorem is the main technical point as regards the successive
study of the Lax-Phillips scattering of generalized Lamb models. It
says that $\A$ is invariant under the flow $U^t_\circ$ and that on
this dense subspace a generalized Lamb model is equivalent to a wave
equation with a high order boundary condition at zero.
\begin{theorem} \label{ponte} 
Let $U_\circ^t$ the strongly continuous group of evolution provided by
Theorem \ref{group}. Then 
$$
U^t_\circ:\A\to\A
$$ 
and
\begin{align*}
&U_\circ^t(\phi_0,M^{-1}v(\phi_0),\dot\phi,M^{-1}v(\dot\phi_0))\\
=&(\phi(t),M^{-1}v(\phi(t)),\dot\phi(t),M^{-1}v(\dot\phi(t)))
\end{align*}
where $\phi(t,x)$ solves the equations 
\begin{equation}\label{cauchy}
\begin{split}
&\partial_{tt}\phi(t,x)=\partial_{xx}\phi(t,x)\,,\qquad x>0\,,\\
&p(\partial_x)\phi\,(t,0_+)=0\\
&\phi(0,x)=\phi_0(x),\quad\dot\phi(0,x)=\dot\phi_0(x)\,.
\end{split}
\end{equation}
Here $p(\partial_x)$ denotes the constant coefficients 
differential operator of degree $2n+1$ $(2n$ if $\theta=0)$ 
associated to the polynomial
$$
p(z)=p_{n+1}(z)-\sum_{i,j=1}^{n}\lambda_i^n\left(V^{-1}\right)_{ij}p_j(z)\,.
$$
\end{theorem}
\begin{proof} Let $(\phi(t),\dot\phi(t))$ be the solution of the 
Cauchy problem 
\begin{equation}\label{cauchy'}
\begin{split}
&\frac{d^2}{dt^2}\,(\phi(t),y(t))=A(\phi(t),y(t))\\
&(\phi(0),y(0))=(\phi_0,y_0)\,,\\
&(\dot\phi(0),\dot y(0))=(\dot\phi_0,\dot y_0)
\end{split}
\end{equation}
with $(\phi_0,y_0,\dot\phi_0,\dot y_0)\in \A$ and let us 
suppose that $(\phi(t),\dot\phi(t))$
is in $\S(\RE_+)$ for all times. Then 
deriving with respect to time the boundary condition $2k$-times, 
$k=0,\dots,n$, using $\ddot\phi(t,0_+)=\phi''(t,0_+)$,
one obtains the $n+1$ equations
\begin{equation*}
\langle w,L^{k}y(t)\rangle=p_{k+1}(\partial_x)\phi(t,0_+)\,,\qquad
k=0,1,\dots n\,.
\end{equation*}
The first $n$ of such equations can be can be rewritten as
$$
My(t)=v(\phi(t))
$$
so that 
$$y(t)=M^{-1}v(\phi(t))\,,\quad
\dot y(t)=M^{-1}v(\dot\phi(t))$$ for all times.
Moreover, inserting the expression for $y(t)$ into the $n$'th
equation one obtains $p(\partial_x)\phi(t,0_+)=0$, so $\phi$ satisfies
\eqref{cauchy}.\par
Conversely let
$\phi(t)$ be the solution of \eqref{cauchy} and put
$$y(t):=M^{-1}v(\phi(t))\,.$$
Then the $n$ equations 
\begin{equation*}
\langle w,L^{k}y(t)\rangle=p_{k+1}(\partial_x)\phi(t,0_+)\,,\qquad
k=0,1,\dots n-1\,.
\end{equation*}
are satisfied. The first equation says that $(\phi,y)$ and
$(\dot\phi,\dot y)$ are in $D(A)$. Deriving each equation two times
with respect
to time and using $\ddot\phi(t,0_+)=\phi''(t,0_+)$ one obtains
\begin{align*}
\langle w,L^{k}\ddot y(t)\rangle=&p_{k+1}(\partial_x)\phi''(t,0_+)
=p_{k+2}(\partial_x)\phi(t,0_+)+\langle w,L^kw\rangle\phi'(t,0_+)\\=&
\langle w,L^{k}(Ly(t)+w\phi'(t,0_+)\rangle
\,,\qquad
k=0,1,\dots n-1\,,
\end{align*}
which implies $\ddot y(t)=Ly(t)+w\phi'(t,0_+)$. 
So $(\phi,y)$ is the solution
of \eqref{cauchy'}. This also show, by unicity, that if the initial
conditions of \eqref{cauchy'} are in $\S(\RE_+)$ then they are in
$\S(\RE_+)$ for all times. This justifies the assumption we made at
the beginning of the proof. 
\end{proof}
The next lemma makes more explicit the polynomial $p$ appearing in 
the previous theorem: 
\begin{lemma}\label{polynomial}  
$$
p(z)=(\theta
z+1)\,\det(z^2-L)-z\sum_{j=1}^{n}\left(\sum_{k=1}^{j}a_{j-k}
\langle w,L^{k-1}w\rangle\right)z^{2(n-j)}\,,
$$
where
$$a_0=1\,,\qquad a_j:=(-1)^{j}
\sum_{i_1<\dots<i_{j}}\lambda_{i_1}\dots\lambda_{i_{j}}\,,\quad
1\le j\le n\,.
$$
\end{lemma}
\begin{proof} 
Put
$$
\tilde a_j:=-\sum_{i=1}^n\lambda_i^n\left(V^{-1}\right)_{ij}\,,\quad 1\le
j\le n\,,\quad \tilde a_{n+1}:=1
$$
and
$$
b_{jk}:=-\sum_{i=1}^{n}\lambda_i^{k-1-j}|w_i|^2\,,\quad 1\le k\le
n+1\,,\ 1\le j\le k-1\,.
$$
By the definitions of $p(z)$, $p_k(z)$, $\tilde a_j$ and
$b_{jk}$ one has
\begin{align*}
p(z)=&\left(z\theta+1\right)\sum_{j=1}^{n+1}\tilde a_jz^{2(j-1)}
+z\left(\sum_{j=1}^{n}b_{j,n+1}z^{2(j-1)}+\sum_{j=2}^{n}\tilde 
a_j\sum_{i=1}^{j-1}b_{ij}z^{2(j-1)}\right)\\
=&\left(z\theta+1\right)\sum_{j=1}^{n+1}\tilde a_jz^{2(j-1)}
+z\sum_{j=1}^{n}\left(\sum_{k=j+1}^{n+1}b_{jk}\tilde a_k\right)z^{2(j-1)}\\
=&\left(z\theta+1\right)p_a(z^2)+z\,p_b(z^2)
\,,
\end{align*}
where
$$
p_a(z)=\sum_{j=1}^{n+1}\tilde a_jz^{j-1}\equiv
\sum_{j=0}^{n}a_jz^{n-j}
\,,$$
$$ 
p_b(z)=\sum_{j=1}^{n}\tilde b_jz^{j-1}\equiv\sum_{j=1}^{n}b_jz^{n-j}\,,
$$
$$
\tilde b_j=-\sum_{k=1}^{n}\sum_{i=j+1}^{n+1}\lambda_k^{i-j-1}
\tilde a_i|w_k|^2=-\sum_{i=j+1}^{n+1}
\tilde a_i\langle w,L^{i-j-1}w\rangle\,.
$$
Since
$$
\sum_{j=1}^{n+1}\tilde a_j\lambda_k^{j-1}
=\lambda_k^n-\sum_{j=1}^{n}\sum_{i=1}^{n}\lambda_i^n\left(V^{-1}\right)_{ij}
V_{jk}=\lambda_k^n-\lambda_k^n=0\,,
$$
the eigenvalues 
$\lambda_1,\dots\lambda_n$ are the roots
of $p_a$. Thus $$
\tilde a_{j}=\sum_{i_1<\dots<i_{n-j+1}}(-1)^{n-j+1}
\lambda_{i_1}\cdots\lambda_{i_{n-j+1}}\,,\quad 1\le j\le n\,,\quad
\tilde a_{n+1}=1\,.
$$
\end{proof}
The next result gives the relation between the roots of $p$ and 
eigenvalues and resonances of the self-adjoint extension $A$:
\begin{lemma}\label{roots} Suppose $\det L\not=0$ and let us define
the couple $(\phi,y)$ by 
$$\phi(x):=e^{zx}\,,\qquad
  y:=M^{-1}v(\phi)\,.
$$ 
Then 
$$p(z)=0\quad\iff\quad
\begin{cases} \phi''=z^2\phi\,,&\\
Ly+w\phi'(0_+)=z^2y\\
\theta\phi'(0_+)+\phi(0_+)=\langle w,y\rangle\,,
\end{cases}
\quad\Longrightarrow\quad
z\in\CO\backslash i\RE\,.
$$
Hence 
$$
p(z)=0\,,\quad\text{\rm Re}(z)\le 0\,,\quad\iff\quad z=-\sqrt\lambda\,,\quad
\lambda\in\sigma_{pp}(A)\cap(0,+\infty)\,.
$$
\end{lemma}
\begin{proof} Since $\phi''=z^2\phi$ implies $z\not=0$ and
$p(0)=(-1)^n
\det L\not=0$, we can take $z\not=0$. By the definition
of $(\phi,y)$ we only need to show that
$$p(z)=0\quad\iff\quad
Ly+wz=z^2y^2\,.
$$
Moreover, beside $\langle w,y\rangle=p_1(z)$, one has 
$$
\langle w, L^ky\rangle=p_{k+1}(z)=z^2p_k(z)-\langle w,L^{k-1}w\rangle\, z
\,,\qquad k=1,\dots,n-1\,.
$$ 
The above equalities, together with $p(z)=0$, gives 
$$
\langle w, L^ny\rangle=p_{n+1}(z)=z^2p_n(z)-\langle w,L^{n-1}w\rangle\, z
$$ 
Thus
$$
\langle w, L^k(Ly+wz-z^2y)\rangle=0\,,\qquad k=0,\dots,n-1\,,
$$ 
i.e. $Ly+wz=z^2y$. Reversing the above argument one has that
$Ly+wz=z^2y$ implies $p(z)=0$.\par 
Suppose now $p(i\nu)=0$, $\nu\in\RE$, so that $Ly+i\nu
w=-\nu^2y$. Since $\langle w,y\rangle=i\nu\theta+1$ we have $$
\langle(L+\nu^2)y,y\rangle=\nu^2\theta-i\nu\,.
$$ 
Since $L+\nu^2$ is symmetric we have $\nu=0$. But $\nu\not=0$ by $\det
L\not=0$.
\end{proof} 
\begin{remark}\label{resonances} By the previuos lemma we have that 
the polynomial $p$
has no purely immaginary roots. The ones in the left half-plane are
real and correspond to eigenvalues, the ones in the right half plane
give rise to not normalizable solutions of the eigenvalue equations
and correspond to resonant states. 
\end{remark}
\end{section}
\begin{section}{Lax-Phillips scattering}
From now on we suppose that
\begin{equation}\label{pp}
\sigma_{pp}(A)=\emptyset\,.
\end{equation}
Here we remark that the successive results hold, with the appropriate
modifications, even without this hypothesis which is just a convenient
choice in order to simplify the exposition.\par
Since we already supposed $\sigma_w(L)=\emptyset$, by Theorem
\ref{spectrum} the above hypothesis means 
$$
\left\{\lambda\in\rho(L)\cap (0,+\infty )\,:\,
\theta+\Gamma(\lambda)=0\right\}=\emptyset\,,
$$
i.e. we are supposing that there are no strictly positive solutions $x$  of the
equation
$$
\frac{1}{x}+\sum_{j=1}^{n}\frac{|w_j|^2}{-\lambda_j+x^2}=\theta\,.
$$
This is true if and only if 
$$
\sigma(L)\subset(-\infty,0)\,,\qquad  \theta\le 0\,. 
$$
Thus hypothesis (\ref{pp}) is satisfied by both 
Examples \ref{lamb} and \ref{acoustic}. 
\par
By Lemma \ref{roots} (\ref{pp}) implies that the polynomial $p$ has no
negative real root and that the complex ones 
(which appear in complex-conjugate pairs since $p$ has real
coefficients) are all contained in the right half-plane.
Thus
$$
\sigma_{pp}(A)=\emptyset\quad\iff\quad \text{\rm Roots$(p)$}
\subset \left\{z\in\CO\,:\,\text{\rm Re$(z)>0$}\right\}\,. 
$$
By functional calculus, since $A$ is injective and negative by
(\ref{pp}) we have that 
$$
U^t_\circ=
\begin{pmatrix} 
\cos t\sqrt{-A}&{\sqrt{-A}}\,^{-1}\sin t\sqrt{-A}\,\\
-\sqrt{-A}\sin t\sqrt{-A}&\cos t\sqrt{-A}\,.
\end{pmatrix}
$$ 
Moreover $U_\circ^t$ extends to a strongly continuous unitary group 
$$U^t:\H\to\H\,,$$ 
where $\H$ is the Hilbert space given by the completion of
$\H_\circ$ with respect to the scalalr product corresponding to energy norm 
$$
\|(\phi,y,\dot\phi,\dot
y)\|_E:=E(\phi,y,\dot\phi,\dot y)^{1/2}\,,
$$
\begin{remark}\label{equi} By Theorem \ref{spectrum} our hypothesis
$\sigma_{pp}(A)=\emptyset$ says that $A$ has purely absolutely
continuous spectrum. Thus 
$$
\lim_{t\to\pm\infty}\,\|y(t)\|=0\,.
$$
Indeed by functional calculus and Riemann-Lebesgue
lemma, denoting by $P(d\lambda)$ the projection-valued measure
corresponding to $A$, one has, for any $(\phi,y,\dot\phi,\dot y)\in \H_\circ$, 
\begin{align*}
&\lim_{t\to\pm\infty}\,
\lambda_iy_i(t)\\=&
\lim_{t\to\pm\infty}\,\langle\sqrt{-A}\,(0,\hat e_i),\sqrt{-A}\cos
t\sqrt{-A}\,(\phi,y)+\sin t\sqrt{-A}\,(\dot\phi,\dot y)
\rangle_{L^2(\RE_+)\oplus\fh}\\
=&\lim_{t\to\pm\infty}\,\int_0^\infty\cos t\sqrt\lambda\,\langle
\sqrt{-A}\,(0,\hat e_i),P(d\lambda)\sqrt{-A}\,(\phi,y)\rangle_{L^2(\RE_+)\oplus\fh} \\
+&\lim_{t\to\pm\infty}\int_0^\infty\sin t\sqrt\lambda\,\langle
\sqrt{-A}\,(0,\hat e_i),P(d\lambda)(\dot\phi,\dot y)\rangle_{L^2(\RE_+)\oplus\fh} =0
\end{align*}
Analogously one has 
$$
\lim_{t\to\pm\infty}\,\|\dot y(t)\|=0\,.
$$
\end{remark}
Define 
$$
\F
:=\left\{(f_-,f_+)\in\S_0(\RE)\times\S_0(\RE)\,:\, p(\partial_x)f_-+
p(-\partial_x)f_+=0\right\}\,,
$$
where
\begin{align*}
\S_0(\RE)
:=&\left\{f\in\S(\RE)\,:\int_{\RE}f(x)\,dx=0\right\}\\
\equiv&\left\{f\in\S(\RE)\,:\, f=g'\,,\ g\in\S(\RE)\right\}\,.
\end{align*}
By considering the Fourier transform (denoted by $\hat{\,}\,$) 
of the differential equation 
\begin{equation}
\label{equaz_diff_gen}
p(\partial_x)f_-+
p(-\partial_x)f_+=0
\end{equation}
one obtains
$$
p(i\kappa)\hat f_-(\kappa)+
p(-i\kappa)\hat f_+(\kappa)=0\,.
$$
Thus one has the following result, which permits to define what will
be the scattering operator.
\begin{lemma}
$$
(f_-,f_+)\in\F\quad\iff\quad
\hat f_+(\kappa)=-\frac{p(i\kappa)}{p(-i\kappa)}\,\hat f_-(\kappa)
\,.
$$
\end{lemma}
\vskip10pt
Since $p$ has real coefficients, we have
$$
\left|\frac{p(i\kappa)}{p(-i\kappa)}\right|=1\,,
$$
therefore 
$$
S_p:\S_0(\RE)\to\S_0(\RE)\,,\qquad (S_pf)\hat{\,}(k)
:=-\frac{p(i\kappa)}{p(-i\kappa)}\,\hat f(\kappa)\,.
$$
extends to an unitary map on $L^2(\RE)$.
\par
\vskip10pt
Let $\phi(t)$ be the solution of (\ref{cauchy}) with initial data
$(\phi,y,\dot\phi,\dot y )\in\A$. Then 
$$
\phi(t,x)=a(x+t)+b(t-x)\,,
$$
where the couple $(a,b)$ is determined on an half-line up
to a constant $c$ by 
\begin{equation}
\label{a}
a(x)=-\frac{1}{2}\int_x^{+\infty}
(\dot\phi(y)+\phi'(y))\,dy+c\,,\quad x\ge 0\,,
\end{equation}
\begin{equation}
\label{b}
b(-x)=\frac{1}{2}\int_x^{+\infty}
(\dot\phi(y)-\phi'(y))\,dy-c\,,\quad x\ge 0\,.
\end{equation}
The functions $a(x)$ and $b(-x)$ are then determined for
the remaining
values of $x<0$ by solving the differential equation 
\begin{equation}
\label{dequation}
p(\partial_x)a+p(-\partial_x)b=0\,.
\end{equation}
The following central result holds:
\begin{theorem}
\label{mainthm}
The map $\I_\circ\equiv(\I_\circ^-,\I_\circ^+)$ defined by
$$
\I_\circ:\A\to\F\,,
\qquad\I_\circ(\phi,y,\dot\phi,\dot y):=(f_-,f_+)\,,\quad f_-=a'\,,\quad f_+=b'\,.
$$
is one-to-one and extends to
an unitary map $$I:\H\to \text{\rm Graph$(S_p)$}\,.$$
\end{theorem} 
\begin{proof}
It is easy to check that the map $\I_\circ:\A\to\F$ is iniective.
It is surjective too. Indeed, if $(f_+,f_-)\in\F$ then
$I_\circ(\phi,y,\dot\phi,\dot y)=(f_+,f_-)$
where
$$
(\phi,y,\dot\phi,\dot y)=
(\phi,M^{-1}v(\phi),\dot\phi,M^{-1}v(\dot\phi))\,,
$$
$$
\phi(x)=a(x)+b(-x)\,,\quad
\dot\phi(x)=a'(x)+b'(-x)\,,
$$
$$
a(x):=\int_{-\infty}^xf_-(y)\,dy\,,\quad
b(x):=\int_{-\infty}^xf_+(y)\,dy\,.
$$
Let us now show that
$$
\|(\phi,y,\dot\phi,\dot y)\|_E^2=
\|I_\circ^-(\phi,y,\dot\phi,\dot y)\|_2^2+\|I_\circ^+(\phi,y,\dot\phi,\dot y)\|_2^2\,.
$$
Since
$$
I_\circ^+=S_pI_\circ^-
$$
and $S_p$ is unitary we need to show that
$$
\|(\phi,y,\dot\phi,\dot y)\|_E^2=2
\|I_\circ^-(\phi,y,\dot\phi,\dot y)\|_2^2\,.
$$
Let $(\phi(t,x),y(t),\dot\phi(t,x),\dot y(t))$ the solution of 
\begin{equation}
\label{sys}
\left\{
\begin{array}{ll}
\ddot\phi(t,x)-\phi''(t,x)=0 \\ 
\ddot y(t)-Ly(t)-w\phi'(t,0_+)=0\\
\theta\phi'(t,0_+)+\phi(t,0_+)-\langle w,y(t)\rangle=0
\end{array}
\right.
\end{equation}
with initial data $(\phi(x),y,\dot\phi(x),\dot y)$. Then, by using the
evolution equation for $y$ and the boundary conditions, one has
\begin{equation}
\label{conto2}
\begin{split}
&\frac{d}{dt}\left(
\|\dot y(t)\|^2-
\langle y(t),Ly(t)\rangle-{\theta}
|\phi'(t,0_+)|^2
\right)\\
=&\dot\phi^*(t,0_+)\phi'(t,0_+)+(\phi')^*(t,0_+)\dot\phi(t,0_+)
\end{split}
\end{equation}
and so
\begin{align*}
&\int_0^\infty
\left(\dot\phi^*(t,0_+)\phi'(t,0_+)+(\phi')^*(t,0_+)\dot\phi(t,0_+)\right)
\, dt\\=&
-\left(\|\dot y\|^2-
\langle y,Ly\rangle-{\theta}|\phi'(0_+)|^2
\right)\,.
\end{align*}
By conservation of energy one has
\begin{align*}
&2E(\phi,y,\dot\phi,\dot y)\\=&\|\dot\phi\|^2_2+\|\phi'\|^2_2+
\|\dot y\|^2
-\langle y,Ly\rangle
-{\theta}|\phi'(0_+)|^2=\\
=&
\int_0^\infty\left(\dot\phi^*(x)\dot\phi(x)+
(\phi')^*(x)\phi'(x)\right)\,dx\\
-&\int_0^\infty\left(\dot\phi^*(t,0_+)\phi'(t,0_+)+
(\phi')^*(t,0_+)\dot\phi(t,0_+)\right)\, dt.
\end{align*}
Inserting in the last equation $\phi(x,t)=a(t+x)+b(t-x)$
and using $b'=S_pa'$, we have
$$
E(\phi,y,\dot\phi,\dot y)=2\|b'\|_2^2=
2\|I_\circ^-(\phi,y,\dot\phi,\dot y)\|^2_2\,.
$$
\end{proof}

We define now the maps $R_\circ^\pm:\A\to\S_0(\RE)$ by
$$
R_\circ^\pm(\phi,y,\dot\phi,\dot y)(x):=I_\circ^\pm(\phi,y,\dot\phi,\dot y)(-x)
$$
and the orthogonal spaces $\H^\pm$ as the closure, with respect to the
energy norm, of 
$$\A^\pm:=\{(\phi,y,\dot\phi,\dot y)\in\A\,:\,
R_\circ^\pm(\phi,y,\dot\phi,\dot y)\in\S^\pm_0(\RE)\}\,,
$$
where
$$
\S^\pm_0(\RE):=\left\{f\in\S_0(\RE)\,:\,f(x)=0\,,\quad \pm
x\le 0\right\}\,.$$
The next theorem show that the subspaces $\H^-$ and $\H^+$ are
incoming and outgoing in the sense of Lax-Phillips scattering theory
(see \cite{[LP]}, \cite{[RS]}, section XI.11). The proof is a straightforward consequence of the previous definitions.
\begin{theorem}
\label{theo:in out}
The subspace
${\mathcal H}^-$ is incoming and the subspace 
${\mathcal H}^+$ is outgoing, i.e.
$$
U^s\H^{-}\subset U^t\H^{-}\,,\ s<t\le 0\,,\quad
\bigcap_{t<0} U^t\H^{-}=\left\{0\right\}\,,\quad
\overline{\bigcup_{t\in\RE}U^t \H^{-}}=\H\,,
$$
$$
U^t\H^{+}\subset  U^s\H^{+}\,,\ t>s\ge 0\,,\quad
\bigcap_{t>0}U^t\H^{+}=\left\{0\right\}\,,\quad
\overline{\bigcup_{t\in\RE}U^t \H^{+}}=\H\,.
$$
\end{theorem}
By the previous theorem and by \cite{[LP]}, chapter II, sections 2 and
3, there follows
\begin{theorem}
\label{theo:sulle R}
The unitary maps
$$
R^\pm:\H\to L^2(\RE)\,,
$$
defined as the closures of the maps $R^\pm_\circ$, 
provide incoming and outgoing translation representations of $U^t$, i.e.
\begin{align*}
&R^\pm  U^t (R^\pm)^{-1}=T^t\,,\\
&R^\pm\H^\pm=L^2(\RE_\pm)\,,\\
&S_{p}^*=R^+ (R^-)^{-1}\,,
\end{align*}
where
$$
T^t:L^2(\RE)\to L^2(\RE)\,,\qquad T^t f(x):=f(x-t)\,.
$$
\end{theorem}
\begin{proof}
The thesis follows from the Theorem \ref{mainthm} and from
simple computations. Otherwise one can use, as we said, the
general theory contained in \cite{[LP]}.
\end{proof}
\end{section}
\begin{section}{The Lax-Phillips semigroup}
We are now interested in the evolution of the states which are neither
incoming in the past nor outgoing in the future. To this end one defines
$$
Z^t:=PU^t P\,,$$
where $P$ is the orthogonal projection onto $$
\K:=\H\ominus(\H^-\oplus\H^{+})\,.$$ Since $ \H^+$ and $\H^-$ are orthogonal 
it is known (see \cite{[LP]}, \cite{[RS]} section XI.11) that $Z^t$ is a
strongly continuous semigroup of contractions on $\K$ for positive times: 
$$
\forall\,t\ge 0\,,\qquad Z^t:\K\to\K\,,
\qquad\|Z^t\|\leq 1\,,\qquad \lim_{t\uparrow\infty}Z^t=0\,.
$$
The next theorem completely characterizes such a semigroup (let us
remark that, by (\ref{pp}), all the roots of $p$ have positive real part):
\begin{theorem} The vector space $\K$ is finite dimensional,$$
\text{\rm dim}\,\K=\text{\rm deg}(p)\,.$$ It is generated by the vectors
\begin{equation}\label{basee}
(R^+)^{-1}\phi_{kj}\,,\qquad k=0,1,\dots,\nu_j-1,\quad
j=1,\dots,m\,,
\end{equation}
$$\phi_{kj}(x)
:=\begin{cases}x^ke^{z_jx}\,,&\text{for $x\le 0$}\\
0&\text{for $x>0$}
\end{cases}\,,$$ 
where  $z_1,\dots z_m$ are the roots of the polynomial $p$ 
and $\nu_1,\dots \nu_m$
the respective multiplicities. Moreover
$$
Z^t=e^{-tB}\,,\qquad\sigma(B)=\left\{z_1,\dots,z_m\right\}\,,
$$
and the matrix representing $B$ with
respect to the basis (\ref{basee}) is the direct sum $B=\oplus_{j=1}^m
B_{j}$ where $B_{j}$ is the $\nu_j\times \nu_j$ matrix
$$
\begin{pmatrix} 
z_j&1&0&\hdots &0\\
0&z_j&2&\hdots&\vdots\\
0&0&z_j&\ddots&0\\
\vdots&\vdots &\ddots&\ddots&\nu_j-1\\
0&0&\dots&0&z_j
\end{pmatrix}\,.
$$
\end{theorem}
\begin{proof}
Since $R_+:\H\to L^2(\RE)$ is unitary, $
R^{\pm}\H^{\pm}=L^2(\RE_{\pm})$ and $R^+(R^-)^{-1}=S_{p}^*$ we have 
$$
R^+\K=L^2(\RE)\ominus\left( 
\left(L^2(\RE_-)\cap S_{p}^*L^2(\RE_-)\right)\oplus L^2(\RE_+)\right)\,.
$$
By our hypotheses $p(i\zeta)\not=0$ for any $\zeta$ in the upper
complex plane $\CO_+$. Thus by
Paley-Wiener theorems the analytic extension to $\CO_+$ of the 
Fourier transform of $f\in \S^-_{0}\cap S_{p}^*\S^-_{0}$
$$
\hat f(\zeta)=-\frac{p(-i\zeta)}{p(i\zeta)}\,\hat g(\zeta)\,,
$$
has no poles and has
zeroes of order $\nu_j$ at $iz_j$, i.e.
$$
\left.\frac{d^k\hat f(\zeta)}{d\zeta^k}\right|_{\zeta=iz_j}
=\frac{(-i)^k}{\sqrt{2\pi}}
\int_{-\infty}^0 x^ke^{z_jx}f(x)\,dx=0\,,\qquad k=0,1,\dots,\nu_j-1\,. 
$$
Thus $$
L^2(\RE_-)\cap S_{p}^*L^2(\RE_-)
=\left\{x^ke^{z_jx}\,,\quad k=0,1,\dots,\nu_j-1,\quad
j=1,\dots,m\right\}^{\perp}
$$
and the finite dimensional subspace $\K$ is whose
generated by vectors \eqref{basee}. These vectors are indipendent
and so the dimension of $\K$ is $\sum_{j=1}^{m}\nu_j=\text{\rm deg}\,p$.
To determine the action of $Z^t$ on $\K$ is enough
to note that the evolution of vectors \eqref{basee} in the
outgoing representation is given by
$$
\phi_{kj}(x-t)\equiv(x-t)^{k}e^{z_j(x-t)}\chi_{(-\infty,0]}(x-t)
$$
Thus
$$\left.\frac{d}{dt}\,\phi_{kj}(x-t)\right|_{t=0}
=-\left(k\phi_{k-1,j}(x)+z_j\phi_{kj}(x\right))
$$
and the proof is done.
\end{proof}

\end{section}

\end{document}